
\magnification=\magstep1
\baselineskip=20pt
\centerline{Two Populations and Models of Gamma Ray Bursts}
\bigskip
\centerline{J. I. Katz}
\centerline{Department of Physics and McDonnell Center for the Space
Sciences}
\centerline{Washington U., St. Louis, Mo. 63130}
\bigskip
\centerline{Abstract}
\medskip
Gamma-ray burst statistics are best explained by a source population at
cosmological distances, while spectroscopy and intensity histories of some
individual bursts imply an origin on Galactic neutron stars.  To resolve this
inconsistency I suggest the presence of two populations, one at cosmological
distances and the other Galactic.  I build on ideas of Shemi and Piran
(1990) and of M\'esz\'aros and Rees (1993) involving the interaction of
fireball debris with surrounding clouds to explain the observed intensity
histories in bursts at cosmological distances.  The distances to the
Galactic population are undetermined because they are too few to affect the
statistics of intensity and direction; I explain them as resulting from
magnetic reconnection in neutron star magnetospheres.
\vfil
\eject
\centerline{1. Introduction}
\medskip
Attempts to explain all the observed gamma-ray bursts (GRB) with a single
population of sources have become progressively more difficult.  On one
hand, their distribution on the sky has been observed, with steadily
improving precision (Atteia, {\it et al.} 1987; Meegan, {\it et al.} 1992),
to be isotropic, an observation which is naturally explained (Usov and
Chibisov 1975; Goodman 1986, Paczy\'nski 1986, Mao and Paczy\'nski 1992,
Fenimore {\it et al.} 1992, Piran 1992) if they are at cosmological
distances.  On the other hand, a number of GRB have been reported to show
(Higdon and Lingenfelter 1990) spectral features at a few tens of KeV and at
about 400 KeV, which are readily interpretable as cyclotron lines and the
two-photon positron annihilation line from the surface of magnetized neutron
stars at Galactic distances, but which are inexplicable at cosmological
distances.  If their validity is accepted, the data appear irreconcilable.

The problem is further complicated by the fact that straightforward models of
radiation transport in GRB at cosmological distances (Goodman 1986;
Paczy\'nski 1986) predict very brief bursts of radiation with thermalized
spectra, in contradiction to observation, while attempts (Brainerd 1992;
Katz 1992) to explain the spatial anisotropy and $\log N$ {\it vs.}~$\log S$
or $V/V_{max}$ distributions of GRB in Galactic models require the assumption
of a spherically symmetric halo of $\sim 100$ Kpc radius.  Finally,
the soft gamma repeaters (SGR) introduce additional confusion.  The fact of
their repetition and the identification of one of them (March 5, 1979; Cline
1980) with a supernova remnant in the LMC point strongly to a Galactic
population, while the presence of spectral features and an 8 second
periodicity (March 5, 1979) indicate origin on a magnetic neutron star.
However, it is unclear that SGR should be considered GRB at all because
their properties, including their spectra, are very different, and arguments
made for SGR may be irrelevant to the problems of GRB.

As a first step towards resolving the apparent inconsistencies, I consider
the obvious possibility that there are two distinct populations of GRB.  A
cosmological population C accounts for most GRB, and explains the statistics
of isotropy and the $\log N$ {\it vs.}~$\log S$ or $V/V_{max}$
distributions.  A population G includes those GRB with spectral lines (a
minority), and originate on neutron stars at Galactic distances.  An individual
GRB cannot be assigned to a population unless it shows spectral lines, but
the majority (probably overwhelming) of those without spectral lines must
be members of population C.  The SGR may be members of population G.  I do
not consider more exotic possibilities, such as GRB arising within the Oort
cloud (Ruderman 1975), because they all seem far-fetched, even though I
myself have discussed one of them (Katz 1993).

In the absence of information about the intensity statistics and angular
distribution of population G alone, it is not possible to discriminate
between a disc and a halo origin.  It will be important
to obtain this information, which may be reducible from archival data.

In \S 2 I discuss a possible mechanism for GRB at cosmological distances,
building on recent suggestions by others.  In \S 3 I more briefly
discuss magnetic reconnection models of GRB at Galactic distances.  The
March 5, 1979 event in the LMC poses an acute problem of gamma-gamma pair
production (Carrigan and Katz 1992) which must be faced, whether or not GRB
of Population G have comparable distances and luminosities, and even if there
is no population G.  \S 4 contains a summary discussion.  Unfortunately,
unambiguous observational tests of the ideas discussed here will not be easy.
\bigskip
\centerline{2. Population C: GRB at Cosmological Distances}
\medskip
The well-known failure of straightforward fireball models to explain the
spectral and temporal properties of GRB led Shemi and Piran (1990) to
consider neutrino-produced fireballs loaded with small (but not zero)
amounts of ordinary matter; they found that the fireball could couple nearly
all of its energy to the matter and (with the right values of the
parameters) could accelerate it to relativistic velocity.  M\'esz\'aros and
Rees (1993) then pointed out that the interaction of this relativistic
debris with surrounding matter might be characterized by times consistent
with the range of GRB rise times and durations (10$^{-3}$--$10^3$ sec).

I build on these ideas.  The environments of GRB's at cosmological distances
are open to much speculation (for example, are they low density galactic
halos or dense nuclei of galaxies?), but the strong clumpiness of interstellar
matter is a consequence of immutable atomic physics (cooling rates), and
isolated discrete clouds are likely under a very wide range of conditions.
The rarity of GRB makes it possible to assume favorable conditions, if these
lie in the range of plausibility; there is no great difficulty if a
considerably larger number of fireballs occurring in less favorable
circumstances do not produce observable GRB.

GRB at cosmological distances require the radiation of $\sim 10^{51}$ erg in
observable gamma-rays.  The complex chain of processes which lead to
gamma-ray emission must be moderately efficient when the parameters do have
favorable values, because the energy radiated as neutrinos by neutron star
collapse, formation, or coalescence is unlikely to exceed $0.3 M_\odot c^2
\approx 5 \times 10^{53}$ erg, and may be considerably less.  Thus,
while we are entitled to assume favorable circumstances to explain the rare
observable GRB, when these circumstances occur the resulting processes must
be reasonably efficient.  The fraction of neutrino energy converted to an
electromagnetic fireball is small.  Efficient conversion requires
neutrino-neutrino collisions at angles in excess of $90^\circ$, but the
neutrinos are generally expanding outward in a neutrino fireball, with
velocity vectors which are tending toward radial outflow.  Optimal head-on
collisions are particularly rare.  The conversion of electromagnetic energy
to particle kinetic energy also has an efficiency $< 1$.  The final conversion
to observable gamma-rays is the hardest part of the problem; it is easy to see
how this could fail entirely.

Relativistic invariants alone limit the amount of kinetic energy available
for radiation by fireball debris.  If a relativistic debris cloud with speed
$\beta_F c$, Lorentz factor $\gamma_F$ and proper mass per unit area $\sigma$
sweeps up a proper mass per unit area $\alpha\sigma$, then the efficiency of
radiation of the debris kinetic energy can be as large as
$$\epsilon = {\alpha \over \alpha + \gamma_F (1 - \beta_F)}. \eqno(1)$$
Values of $\epsilon > 1/2$ are obtained for $\alpha > 1/[\gamma_F(1 +
\beta_F)] \approx 1/(2\gamma_F)$.  Collisions with a very broad range of
clouds of circum-fireball matter are consistent with efficient conversion of
kinetic energy to radiation.  This is fortunate, because efficient
production of GRB requires that common circum-fireball environments produce
observable GRB in most directions; it is not possible to insist on
fortuitous geometries or on special values of the parameters.

Most of the kinetic energy will become available when the debris has swept
up only a very little matter.  If the energy of the explosion is $Y$ then
the proper mass of debris is $Y/\gamma_F c^2$, and half the kinetic energy
will become available when the swept-up proper mass is $Y/2\gamma_F^2 c^2$.
In a uniform medium of density $\rho$ this will occur at an interaction radius
$$r_I = \left({3 Y \over 8 \pi \gamma_F^2 c^2 \rho}\right)^{1/3} \sim
2 \times 10^{15}\ {\rm cm}, \eqno(2)$$
where the numerical estimate assumed $Y = 10^{52}$ erg, $\rho = 10^{-24}$
g/cm$^3$ and $\gamma_F = 10^4$.

The hardest part of the problem is turning the kinetic energy of the
relativistic debris into the observed gamma-rays.  The collision length of
relativistic protons in ordinary matter is about 50 g/cm$^2$, or about 100
Mpc at typical interstellar densities.  Clearly, some collective process is
necessary, and it must couple the proton and ion energy into that of
electrons, which radiate more readily.  Even relativistic electrons do not
radiate rapidly under interstellar conditions; the radiation length of a
$10^{13}$ eV electron (corresponding to equipartition with a $\gamma_F \sim
10^4$ proton) for Compton scattering on a 3$^\circ$K black body radiation
field is $\sim 10^{23}$ cm, excessive by many orders of magnitude.

In order to obtain short pulses of radiation at distances of order those
given by (2) it is necessary that a coherent relativistically expanding
front of radiating particles be directed nearly toward the observer.
It is not sufficient that individual
particles be observed only when directed towards the observer, a condition
met by most relativistic radiation processes.  Therefore, ambient magnetic
fields must not deflect particles significantly from their initial spherical
expansion.  This condition will be satisfied if the magnetic energy
$E_{mag}$ in the interaction sphere of radius $r_I$ is very much less than
the debris energy, so the debris can sweep away the ambient magnetic field
without significant deflection.  If equipartition is assumed between the
ambient magnetic field $B_{ISM}$ and an ambient turbulent velocity field
$v_{ISM}$, then $E_{mag}/Y \sim v_{ISM}^2 / \gamma_F^2 c^2 \ll
1/\gamma_F^2$, so the ambient field may safely be ignored.
\medskip
\noindent
a) Shock Structure
\smallskip
When the debris shell collides with a cloud of ambient matter the resulting
flow may be complex.  If the shell and the cloud each initially had uniform
density and velocity and negligible (on a relativistic scale) temperature,
the geometry is slab-symmetric, and all bulk velocities are normal to the
planes of symmetry, then the resulting shock structure is shown in Figure 1.
There are two shocks S1 and S2 and, in general, a contact discontinuity CD
separating shocked fireball debris from shocked cloud.  The equations
relating the conditions in the four regions are cumbersome, except in the
special symmetric case in which debris and cloud initially had the same
composition and proper density.  In this case, which I assume, there is no
contact discontinuity and conditions in regions 2 and 3 are identical, as
are those in regions 1 and 4.

The relativistic shock conditions (Landau and Lifshitz 1959) may be used to
determine physical conditions.  The thermodynamic variables are the proper
internal energy density $e$, the proper pressure $p$, and the proper
enthalpy density $w = e + p$.  In the unshocked cloud $e_1 = n_1 m_a c^2$,
where $n_1$ is the proper atomic number density and $m_a$ is the proper mass
per atom, and $p_1 = 0$.  In the shocked cloud $p_2 = e_2/3 \gg e_1$ in the
extreme-relativistic (ER) limit.  I shall refer to the frame of the
unshocked interstellar material as the local observer's frame; transformation
to our frame requires application of the cosmological redshift.  Then,
to lowest nontrivial order in $e_1/e_2 \ll 1$, the velocities of the fluids
with respect to the shock front S1 are
$${v_1 \over c} = \left[{(p_2 - p_1)(e_2 + p_1) \over (e_2 - e_1)(e_1 +
p_2)}\right]^{1/2} \approx 1 - {e_1 \over e_2} \eqno(3)$$
and\footnote*{Note the assertion in the first edition of Landau and
Lifshitz that in the ER limit $v_2 \to c/3^{1/2}$ is a typographical error;
the correct limit $v_2 \to c/3$ is given in later editions.}
$${v_2 \over c} = \left[{(p_2 - p_1)(e_1 + p_2) \over (e_2 - e_1)(e_2 +
p_1)} \right]^{1/2} \approx {1 \over 3} \left(1 + 2{e_1 \over e_2}\right).
\eqno(4)$$
The velocity discontinuity $v_{12}$ between fluids 1 and 2, measured in the
frame of either, is obtained from the relativistic expression for the
subtraction of velocities
$${v_{12} \over c} = {v_1/c - v_2/c \over 1 - v_1 v_2/c^2} \approx 1 - 2{e_1
\over e_2}; \eqno(5)$$
this velocity is also the velocity $v_{2L}$ of shocked fluid 2 in the local
observer's frame.  The velocity $v_1$ is also the speed of the shock S1 in
that frame.

Fluids 2 and 3 have the same velocity and, given our assumptions that $n_1 =
n_4$ and $e_1 = e_4$, the same values of the thermodynamic variables.  Then
the velocity of fluid 3 in the frame of shock S2 is $-v_2$, and the speed
of fluid 4 in that same frame is $-v_1$.  The expressions for combinations
of relativistic velocities may then be used to obtain the following results
in the local observer's frame:
$${v_3 \over c} = {v_{12} \over c} \approx 1 - 2{e_1 \over e_2}, \eqno(6)$$
$${v_{S2} \over c} \approx 1 - 4{e_1 \over e_2}, \eqno(7)$$
$${v_4 \over c} \approx 1 - 2\left({e_1 \over e_2}\right)^2. \eqno(8)$$

It is now possible to calculate $e_2$ from the debris Lorentz factor
$\gamma_F$, defined in the local observer's frame, using Equation 8:
$$\gamma_F \equiv {1 \over (1 - (v_4 / c)^2)^{1/2}} \approx {e_2 \over
2 e_1}, \eqno(9)$$
$$e_2 \approx 2 n_1 m_a c^2 \gamma_F. \eqno(10)$$
The Lorentz factor $\gamma_{2L}$ of the shocked material in the local
observer's frame is obtained from $v_{2L} = v_{12}$, Equations 5 and 10:
$$\gamma_{2L} \approx \gamma_F^{1/2}. \eqno(11)$$

The detailed mechanics of the shock are obscure, but must be collisionless
in order to form a shock at all.  The shocked matter need not be in
thermodynamic equilibrium.  Heating of the shocked material by plasma
instabilities is the source of dissipation; the distribution functions of
particle energies will not be (relativistic) Maxwellians, but are more
likely to be power laws.  The distribution of energy between electrons and
ions in uncertain.  In a highly relativistic shock, as we expect here,
electrons and ions are kinematically very similar (identical in the ER
limit), so I will assume that the distribution of particle energies is
independent of species; roughly half the post-shock energy resides in
electrons.  Any neutral matter does not interact with the shock, so that
the density $n_1$ refers only to the ionized component.  At distances $\sim
r_I$ (Equation 2) the interstellar material will largely have been ionized by
the flash of radiation associated with the fireball or by collision with
debris.

Using the shock jump conditions for the proper enthalpy $w$, $w_1 = n_1 m_a
c^2$, and the ER limit $w_2 \approx 4 e_2/3$ yields the proper atomic
density
$$n_2 \approx 2 \left({n_1 e_2 \over m_a c^2}\right)^{1/2}. \eqno(12)$$
Define $\gamma_2$ by the relation
$$\gamma_2 \equiv {e_2 \over n_2 m_a c^2}; \eqno(13)$$
then the mean energy (in the frame of fluid 2) per particle is $\gamma_2 \mu
m_p c^2$, where $\mu m_p$ is the mean proper mass per particle.  For pure
ionized hydrogen $\mu = 0.5$, while for the usual cosmic abundances (fully
ionized) $\mu = 0.62$.  The mean Lorentz factor of an electron (in the
frame of fluid 2) is
$$\gamma_{2e} = \gamma_2 {\mu m_p \over m_e}. \eqno(14)$$

The proper density $n_2$ (Equation 12) may be rewritten, using Equation 13,
as
$$n_2 \approx 4 n_1 \gamma_2, \eqno(15)$$
reproducing the result $n_2 = 4 n_1$ for a strong but nonrelativistic shock
($\gamma_2 \to 1$) in a gas with an adiabatic exponent of 5/3.  The Lorentz
factor $\gamma_2$ is found from its definition (Equation 13), and Equations
9 and 15:
$$\gamma_2 \approx {1 \over 2}\left({e_2 \over e_1}\right)^{1/2} \approx
\left({\gamma_F \over 2}\right)^{1/2}. \eqno(16)$$
In the local observer's frame most of the particles are narrowly collimated
in the direction of the motion of fluid 2, and the typical Lorentz factor is
larger than those given in Equations 14 and 16 by a factor $\sim
\gamma_{2L}$ (Equation 11).  The angular width of collimation depends on the
angular distribution of the particle momenta in the proper frame of fluid 2,
which is unknown.  If this is isotropic, then the locally observed angular
width (for electrons as well as ions)
$$\theta_0 \sim \gamma_{2L}^{-1} \approx \gamma_F^{-1/2}. \eqno(17)$$
Note that this is a much broader angular distribution than the locally
observed radiation pattern from a single particle, whose Lorentz factor is
$\sim \gamma_F$ (ions) or $\sim \gamma_F m_p/m_e$ (electrons).
\medskip
\noindent
b) Time Dependence
\smallskip
The geometry of radiation from an advancing spherical shock is shown in
Figure 2.  The distant (but cosmologically local) observer may first see a
flash of radiation from the fireball itself, whose arrival time is taken as
$t = 0$.  If the fireball is a consequence of the merger of binary neutron
stars, as often assumed, the initial pulse includes bursts of neutrino and
gravitational radiation, as well as electromagnetic radiation.  Their
emission is essentially simultaneous, although their arrival may be affected
by dispersion arising from plasma refraction, neutrino rest mass (if they
have any), {\it etc.}  The initial electromagnetic flash is expected to have
a thermal spectrum and to be extremely brief ($\ll 10^{-4}$ sec) because of
the small size of the fireball (Goodman 1986; Carrigan and Katz 1992); if,
as assumed here, the fireball energy is largely converted (Shemi and Piran
1990) to kinetic energy of debris this initial flash may be unobservably
faint.   However, if it is observed the time interval between it and the
rest of the GRB is an important constraint on the emission geometry.

Radiation emitted from a point $(r,\theta)$ on the expanding spherical shell
arrives at the observer at a time
$$\eqalign{t&\approx {r(1 - \cos\theta) \over c} + r\left({1 \over v} - {1
\over c}\right)\cr &\approx {r(1 - \cos\theta) \over c} + {r \over u},\cr}
\eqno(18)$$
where $v$ is the shell's expansion velocity and the parameter (dimensionally
but not physically a velocity) $u \equiv vc/(c - v)$.  If the angular
distribution of radiated intensity, measured in the local observer's frame,
is $f(\theta^\prime)$, where $\theta^\prime$ is the angle from the normal to
the radiating surface, then the energy $dE$ radiated by a patch of area $dA$
is
$$dE = f(\theta^\prime)\, dA. \eqno(19)$$
Radiation directed toward the observer has $\theta^\prime = \theta$.  Using
$dA = 2 \pi r^2 \sin\theta\, d\theta$ and $dt = r \sin\theta\, d\theta/c$
yields the observed power
$$P(t,r) = {dE \over dt} = 2 \pi c r f(\theta). \eqno(20)$$

The function $f(\theta)$ is proportional to a convolution of the angular
distribution of the radiating particles and their radiation pattern; as
previously discussed, the latter is expected to be narrower than the former.
A plausible guess is then
$$f(\theta) \propto {1 \over \theta^2 + \theta_0^2}, \eqno(21)$$
where the angular width $\theta_0 \sim \gamma_F^{-1/2}$ is essentially the
same as that of the momentum distribution (Equation 17).  The pulse shape is
then obtained from Equations 20 and 21, using Equation 18 to eliminate
$\theta$:
$$P(t,r) \propto \cases{{\displaystyle r \over \left[\displaystyle\left(t-{r
\over u}\right){2c \over r} + \theta_0^2\right]}&$t \ge {\displaystyle r
\over\displaystyle u}$\cr 0,&$t < {\displaystyle r \over\displaystyle u}$\cr}
\eqno(22)$$
where the approximation $\cos\theta \approx 1 - \theta^2/2$ has been used.
$P(t,r)$ has been plotted in Figure 3, where the dimensionless parameter
$\tau \equiv 2ct/r\theta_0^2$ has been defined.

The pulse form of Equation 22 should be regarded only as an envelope, for
the actual pulse shape will be modulated by the spatial distribution of
matter which the debris shell sweeps up.  One striking feature of Equation
22 is its abrupt rise, consistent with the observed rapid rises of some
GRB.  The characteristic width of this function is
$$\Delta t \sim {r \theta_0^2 \over 2c} \sim {r \over \gamma_F c}.
\eqno(23)$$
Use of $r \sim r_I$ (Equation 2) and $\gamma_F \sim 10^4$ yields $\Delta t
\sim 10$ sec, the right order of magnitude for the duration of GRB.  Much
longer or shorter $\Delta t$ may be possible for plausibly different values
of the parameters, particularly the cloud density, which is uncertain even
to order of magnitude.

The debris shell and shock propagate into a very heterogeneous medium.  The
effects of structure in $\theta$ are shown by the dashed lines in Figure 3,
which assume a cloud uniform in the range $\theta_1 \le \theta \le
\theta_2$, with abrupt boundaries.  A more realistic gradual density profile
or shape would produce a gradual rise and decay; the abrupt rise remains if
(and only if) the cloud includes the line $\theta = 0$.  Thus this abrupt
rise is expected for some, but perhaps not all, GRB, in accord with
observations.

A complete intensity profile of a GRB requires the integration of Equation
(22) over $r$:
$${\cal P}(t) = \int_0^{ut} P(t,r) g(r)\, dr, \eqno(24)$$
where the weighting function $g(r)$ includes both the fact that the energy
available for radiation falls off for $r > r_I$ and the effects of
clumpiness of the ambient matter as a function of $r$.  The observable
region is a narrow half-cone of apical angle $\sim \theta_0$: unless the
scale of spatial structure is $< r_I \theta_0 \sim 2 \times 10^{13}$ cm,
clumpiness is more likely to be apparent as a function of $r$ than of
$\theta$, justifying the use here of Equation 22 which ignored any
dependence of density on $\theta$.

In the absence of spatial heterogeneity $g(r)$ may be taken to impose a
cutoff at $r \approx r_I$, so that
$${\cal P}(t) \approx \int_0^{\min(r_I,ut)} P(t,r)\, dr \propto
\int_0^{\min(r_I,ut)} {r^2\, dr \over 2ct + r\left(-{2c \over u} +
\theta_0^2\right)}. \eqno(25)$$
The integral is elementary, but cumbersome, and of limited quantitative
interest because of the artificiality of the assumed uniform density; the
rise time is $r_I/u$.  Two possibilities should be distinguished:
\item{1.}For a shock S1 propagating through a homogeneous medium $v/c
\approx 1 - e_1/e_2 \approx 1 - 1/(2\gamma_F)$ and $u \approx 2 \gamma_F c$.
Then, for $r_I$ given by Equation 2 and $\gamma_F = 10^4$, the rise time is
several seconds, given by Equation 23 and inconsistent with very rapid rise
times.  This corresponds to a long GRB pulse envelope.
\item{2.} On the other hand, the fireball debris propagates through a vacuum
or very low density intercloud medium with a speed $v/c = (1 -
1/\gamma_F^2)^{1/2}$, so that $u \approx 2 \gamma_F^2 c$; impacts upon small
discrete clouds scattered within a region of size $\sim r_I$ will only
introduce a time-width $\sim 10^{-3}$ sec, or less.  This might not
measurably broaden the abrupt rise given by Equation 22 and in Figure 3.

Integration over $r$ introduces two broadenings, one $O(r_I/2\gamma_F^2 c)$
associated with the entire emission region of size $\sim r_I$, where the
low intercloud density is appropriate, and another $O(r_c/2\gamma_F c)$
associated with
individual clouds of size $r_c \ll r_I$.  These broadenings may each be much
less than the envelope width (Equation 23), permitting an observed signal
resembling that of Figure 3.  Several sub-pulses may be observed if the
debris shell collides with several isolated clouds in a medium dense enough
to slow the intercloud shock, so that interpulse times are $O(r_I/2\gamma_F
c)$.

The actual situation is much more complicated than can be discussed here.
For example, debris shell and clouds need not have the same proper
densities, and each is likely to be spatially heterogeneous.  Shocks
propagating through heterogeneous media vary their strength, and produce
associated continuous rarefactions and compressions.  It is plausible that
the complex structure of observed GRB could be explained by the interaction
of relativistic debris shells with clumpy media, but more quantitative
results would require numerical relativistic hydrodynamic calculations.
\medskip
\noindent
c) Radiation
\smallskip
The hardest part of the problem is turning electron kinetic energy into the
observed radiation.  Even though the typical Lorentz factor of an electron
in the local observer's frame is $\sim \gamma_{2L}\gamma_{2e} \sim \mu
\gamma_F m_p/m_e$ (from Equations 11, 14, and 16), their rate of synchrotron
radiation and Compton scattering in plausible interstellar magnetic and
radiation fields is low.  I therefore make the radical suggestion that the
collisionless shock produces approximate equipartition between the magnetic
energy density and the particle energy density.  The proper magnetic field
$B_2$ and energy density in fluid 2 are then, using Equation 10,
$${B_2^2 \over 8 \pi} = \zeta e_2 \approx 2 \zeta n_1 \gamma_F m_a c^2,
\eqno(26)$$
where $\zeta \le 1/2$ is a phenomenological parameter describing the
approach to equipartition.  The synchrotron energy loss time for an electron
with Lorentz factor given by Equation 14, assuming no correlation between
the direction of the electron momentum and that of the magnetic field,
is then, taking pure hydrogen composition
$$t_{r2} \approx \zeta^{-1} \left({n_1 \over 1\, {\rm cm}^{-3}}\right)^{-1}
\gamma_F^{-3/2}\ 1.1 \times 10^7\ {\rm sec}. \eqno(27)$$
The radiating volume is moving toward the observer with a bulk Lorentz
factor $\gamma_{2L}$ (Equation 11), so that application of a Lorentz
transform yields the local observer's measured radiation time
$$\eqalign{t_{obs} &=\gamma_{2L}(t_{r2} - v_{2L}^2 t_{r2}/c^2)\cr &=
t_{r2}/\gamma_{2L}\cr &\approx \zeta^{-1} \left({n_1 \over 1\, {\rm cm}^{-3}}
\right)^{-1} \gamma_F^{-2}\ 1.1 \times 10^7\ {\rm sec}.} \eqno(28)$$
For a plausible interstellar cloud density $n_1 > 1$ cm$^{-3}$ and $\gamma_F
\approx 10^5$, $t_{obs}$ may be a millisecond or less.  This justifies the
assumption, made implicitly in the discussion of GRB rise times and pulse
lengths, that shock-accelerated electrons radiate instantaneously; properly,
the pulse profiles predicted by Equations 24 and 25 should be convolved with
a broadening function which includes the radiation time, and which has a
width $t_{obs}$ in the local observer's frame.

The characteristic frequency of synchrotron radiation, measured in the frame
of fluid 2, is obtained from standard expressions using Equations 14 and 26.
For pure hydrogen the result is
$$\nu_2 \sim \zeta ^{1/2} \left({n_1 \over 1\, {\rm cm}^{-3}}\right)^{1/2}
\gamma_F^{3/2}\ 3 \times 10^{11}\ {\rm sec}^{-1}, \eqno(29)$$
while Lorentz transformation to the local observer's frame, using Equation
11, yields
$$\nu_{obs} \sim \zeta^{1/2} \left({n_1 \over 1\, {\rm cm}^{-3}}\right)^{1/2}
\gamma_F^2\ 3 \times 10^{11}\ {\rm sec}^{-1}. \eqno(30)$$
MeV photons may be observed for $\zeta = 1/2$ if $n_1 \sim 1$ cm$^{-3}$ and
$\gamma_F \sim 4 \times 10^4$, for example.

It has also been observed (Fishman 1993) that many GRB, or subpulses within
them, show a progressive spectral softening with time.  This is qualitatively
explained using Figure 2 and Equation 18.  If the radiation field is
isotropic in the frame of fluid 2 (as will be the case if the particle
distribution and magnetic field directions are isotropic) and has a
characteristic photon energy, then in the local observer's frame the
spectral hardness above this characteristic spectral peak will be a
decreasing function of $\theta$.  Higher frequency photons are
preferentially observed from smaller values of $\theta$, which arrive
earlier in the burst or sub-pulse, while lower frequency photons are observed
over a wider range of $\theta$ and hence over a longer time.  A quantitative
prediction for the spectral evolution with time could be made by numerical
integration of the synchrotron emission function, but would depend on
(uncertain) assumptions made regarding the energy and angular distributions
of the radiating electrons.
\bigskip
\centerline{3. Population G: Galactic GRB}
\medskip
GRB which show spectral features, typically around a few tens of KeV and at
400 KeV, have long been identified with Galactic magnetic neutron stars and
are inexplicable at cosmological distances.  Their distances cannot be
determined from available data, and could be $< 100$ pc, $\sim 100$ Kpc, or
anything in between.  The familiar arguments concerning the mechanisms of
GRB at Galactic distances center on two issues: the source of energy, and
the physical conditions in the emitting region.  The problems are harder,
the greater the assumed distances.  The observation of the March 5, 1979
event at a likely distance of 55 Kpc (Cline 1980) forces the consideration
of distances of that order, and of correspondingly high luminosities, even
though it is unclear whether it (a SGR) was a member of the Population G of
GRB, or represented a distinct third class of objects.

The central problem of distant and luminous gamma-ray sources is gamma-gamma
pair production (Cavallo and Rees 1978; Schmidt 1978; Katz 1982; Epstein 1985;
Carrigan and Katz 1992).  This process does not permit the escape of a large
luminosity of MeV gamma-rays from a small region unless they are collimated,
and thus excludes many models of GRB or SGR at Galactic halo or
cosmological distances.  It is well known that this problem is avoided
in a collimated relativistic outflow of radiating matter, a
consideration which led to the popularity of fireball models, in which an
opaque cloud of radiation and pair gas adiabatically expands and cools until
its particles' velocity vectors are collimated outward.  However, fireballs
are conspicuously incapable of producing low redshift (400 KeV) annihilation
lines, line features at tens of KeV, or the observed long and complex time
structure.  The interaction of fireballs with their environment may solve
the temporal problem, as discussed in \S 2, but offers no hope of solving the
spectral problem.  The case for magnetic neutron stars for GRB or SGR with
spectral lines remains as strong as the data.

It is usually assumed that the radiating region of a GRB or SGR in a neutron
star model is dominated by pair plasma, with $n_+ \gg n_i$, where $n_+$ and
$n_i$ are the positron and ion densities, respectively.  This assumption is
made, in analogy to fireballs, even for non-fireball models, perhaps because
the threatened gamma-gamma pair production catastrophe seems a likely source
of dense pair plasma, and because the observed annihilation line requires a
source of positrons.  However, the assumption of pair dominance may not be
justified in non-fireball models, such as are required to explain Population
G GRB.  If sufficient gamma-ray collimation is present to avoid a
gamma-gamma pair production catastrophe, then the production of pairs may be
negligibly small.  When the observation of annihilation radiation provides
empirical evidence for the production of some positrons, it should be
remembered that an observably narrow annihilation line requires temperatures
$< 50$ KeV, and may be produced by a comparatively small number of positrons
precipitated onto the cool neutron star surface; a hot pair plasma does not
produce a recognizable annihilation line.

If a pair plasma is not an expanding fireball, it must be trapped on
magnetic field lines (Carrigan and Katz 1992).  Gravitation is unimportant
for pairs, so they fill a magnetosphere (presumably of a magnetic neutron
star).  However, they quickly (in a free-flight time) precipitate onto the
stellar surface, where they annihilate, because they more rapidly
radiate their transverse momentum by the cyclotron process; even if the
radiation density is sufficient to maintain most leptons in excited Landau
(magnetic) states (a condition satisfied under only the most extreme
conditions), their interaction with the radiation field destroys their
transverse adiabatic invariant.  In this magnetosphere-filling geometry the
emergent radiation, by whatever process, is not collimated, and gamma-gamma
pair production imposes its usual limits on the emergent flux of MeV
gamma-rays.

It may be more satisfactory to consider an electron-ion plasma with only a
small admixture of positrons (sufficient to produce the observed
annihilation line), so that $n_+ \ll n_i$.  An electric field may accelerate
the electrons, which radiate by bremsstrahlung or by the cyclotron process
after elastic scattering on the ions raises them to excited Landau states.
Because most of the leptons are negative, they form a broadly collimated
beam; if they are relativistic the resulting radiation is similarly
collimated and there is no gamma-gamma pair production catastrophe or limit
(other than a Planck function at an effective temperature characterizing
the electron distribution function) on the emergent intensity.

In contrast, an electric field acting on a pair gas heats it but imparts no
net momentum to the leptons; two counterstreaming beams of gamma-rays
readily produce pairs, rapidly achieving equilibrium with them and limiting
the emergent intensity.  A minority admixture of positrons in an
electron-ion plasma produces only a proportionately small countercurrent of
gamma-rays to those produced by the electrons.  This countercurrent removes
an equal current of electron-produced gamma-rays by pair production, but the
remaining electron-produced gamma-rays form a collimated beam and escape
comparatively freely, suffering little or no (depending on the degree of
collimation) gamma-gamma pair production.  This is a consequence of the net
momentum imparted to the lepton-photon system by the electric field, in
analogy to the momentum imparted to a sector of a fireball by adiabatic
expansion.

It is possible to make simple rough estimates of the parameters of an
electrically heated ion-electron sheet plasma, which might be the source
region of a GRB of Population G, following Katz (1993).  Consider a sheet of
thickness $L$, be composed of positive ions of charge $Z$ and density
$n_i$ ($n_e = Z n_i$), have transverse optical depth $\tau$ and temperature
$T$, and radiate power per unit area $P$.  Define the dimensionless power
per unit area $p \equiv P \hbar^3/(m_e^4 c^6)$, dimensionless thickness
$\ell \equiv L m_e c^2/e^2$ and temperature $t \equiv k_B T/(m_e c^2)$,
where these quantities have been scaled to values characteristic of a
relativistic electron (or pair) gas.  The characteristic radiant
intensity $m_e^4 c^6/\hbar^3 = 4.3 \times 10^{35}$ erg/(cm$^2$ sec) and
length $e^2/(m_e c^2) = 2.8 \times 10^{-13}$ cm (the classical electron
radius).  The optical depth is
$$\tau \sim n_e \sigma_0 L, \eqno(31)$$
where the characteristic cross-section $\sigma_0 \equiv e^4/(m_e^2 c^4)$.
At nonrelativistic energies the appropriate cross-section is the Thomson
cross-section $8 \pi \sigma_0/3$, but at semi-relativistic energies of
interest $\sigma_0$ may be a fair approximation.  The observation of a
nonthermal spectrum implies that $\tau$ cannot much exceed unity, but $\tau
\sim 1$ and $\tau \ll 1$ are each possible.

The large field of a magnetized neutron star has a number of effects.
It enters the argument of the effective Coulomb logarithm in collisional
processes, typically reducing it to $\ln\Lambda \approx \ln(k_B T m_e c /
(\hbar e B))$ (Katz 1982).  Both bulk motion and current flow are restricted
to be parallel to the field lines, justifying the assumption of thin sheet
geometry and making the field distribution nearly force free (${\vec J}
\times {\vec B} = 0$).  Perhaps most important, it means that any electron
energy resulting from motion perpendicular to the field is immediately
radiated.  Even in conditions characteristic of the March 5, 1979 event at
55 Kpc the radiation density is far below Planckian (for $t \sim 1$), so
that electrons may be assumed to be in their ground magnetic state until
collisionally excited, and then to radiate as if in vacuum.  The radiation
rate per unit area may therefore be estimated using standard expressions for
elastic scattering:
$$\eqalign{P&\sim n_e n_i k_B T \left({k_B T \over m_e}\right)^{1/2}
\sigma_0 Z^2 \left({m_e c^2 \over k_B T}\right)^2 L\cr &\sim {n_e^2 e^4
\over m_e c}{LZ \ln\Lambda \over t^{1/2}}.\cr} \eqno(32)$$
This expression may be inverted, using the definitions of $p$, $\ell$, $t$,
and $\alpha \equiv e^2/(\hbar c)$, to give
$$n_e \sim \left({p t^{1/2} \over \ell \alpha^3 Z \ln\Lambda}\right)^{1/2}
{m_e^3 c^3 \over \hbar^3}; \eqno(33)$$
the characteristic density $m_e^3 c^3/\hbar^3 = 1.8 \times 10^{31}$
cm$^{-3}$.  An alternative expression for $n_e$ is obtained from Equation 31:
$$n_e \sim {\tau \over \ell \alpha^3}{m_e^3 c^3 \over \hbar^3}. \eqno(34)$$
Equating the expressions (33) and (34) yields a result for the thickness:
$$\ell \sim {\tau^2 Z \ln\Lambda \over p t^{1/2} \alpha^3}. \eqno(35)$$
Very roughly, $p \sim 10^{-4}$ for the March 5, 1979 event, if in the LMC,
and $p \sim 10^{-7}$ for a typical observed GRB if at 100 Kpc distance,
corresponding to $L \sim 1.4 \tau^2$ cm and $L \sim 1.4 \times 10^3 \tau^2$
cm, respectively, where $Z = 26$, $t = 0.1$, and $\Lambda = 10$ were
taken.  The densities are correspondingly high.

If the radiating sheet is driven by magnetic reconnection, as is plausible,
then the radiated power may be related to the electrical work done:
$$P \sim \sigma_{el} E^2 L, \eqno(36)$$
where $E$ is the electric field (properly, its component parallel to $\vec
B$) and a nonrelativistic expression (Spitzer 1962) is used for the (cgs)
electrical conductivity:
$$\eqalign{\sigma_{el}&\approx 2 \left({2 \over \pi}\right)^{3/2}{(k_B
T)^{3/2} \over m_e^{1/2} e^2 \xi Z \ln\Lambda}\cr &\approx {t^{3/2} \over
\xi Z \ln\Lambda}{m_e c^3 \over e^2},\cr} \eqno(37)$$
where the parameter $\xi \ge 1$ is a correction factor which allows
for the possibility of anomalous (plasma instability) resistivity when
current densities and electron drift velocities are large and for the
decrease in conductivity when electron velocities approach $c$.  The
characteristic conductivity $m_e c^3/e^2 = 1.1 \times 10^{23}$ sec$^{-1}$.
Equations (36) and (37) may be combined with the definitions
of the various dimensionless parameters to give the electric field, current
density, and electron drift velocity:
$$E \sim \left({p \alpha^3 \xi Z \ln\Lambda \over \tau t^{3/2}} n_e m_e
c^2\right)^{1/2}; \eqno(38)$$
$$j = \sigma_{el}E \sim {p t \alpha^3 \over \tau \xi^{1/2} \ln\Lambda}
{m_e^3 c^7 \over e^5}; \eqno(39)$$
$$v_{dr} = {j \over n_e e} \sim {c t^{1/2} \over \xi^{1/2}} \sim \left({k_B
T \over m_e}\right)^{1/2}{1 \over \xi^{1/2}}. \eqno(40)$$
The characteristic current density $m_e^3 c^7/e^5 = 6.5 \times 10^{38}$
esu/(cm$^2$ sec).  The drift velocity is thus comparable to the electron
thermal velocity unless $\xi \gg 1$; ion-acoustic instability is likely
unless $\xi > m_i/(Zm_e)$.

GRB mechanisms based on magnetic reconnection, as in Solar flares, have
been qualitatively discussed for many years (Ruderman 1975).  They
may explain GRB energetics and phenomenology (Katz 1982).  Magnetic
reconnection by sheet currents provides a natural explanation of the
electrically heated sheets discussed in this section.  If this model is
assumed, then Maxwell's equations provide an additional relation among $j$,
$L$, and the magnetic field $B$, permitting further constraints to be placed
on the parameters.  Because $\vec j$ is parallel to $\vec B$, the direction
of $\vec B$ rotates across a sheet current without changing its magnitude
$B$.  If the total angle of rotation across a uniform current sheet is
$\pi$, then
$$L = {Bc \over 4j}. \eqno(41)$$
Defining the usual characteristic magnetic field $B_c \equiv m_e^2 c^3
/(e \hbar) = 4.4 \times 10^{13}$ gauss and the dimensionless parameter $b
\equiv B/(4B_c)$, Equation 41 may be rewritten
$$\ell \sim {\tau b \xi^{1/2} Z \ln\Lambda \over p t \alpha^2}. \eqno(42)$$
Equating this expression to equation (35) yields
$$\tau \sim {b \alpha \xi^{1/2} \over t^{1/2}}. \eqno(43)$$
For plausible $t \sim 1$ and $b \sim 0.02$, $\tau \sim 10^{-4} \xi^{1/2}$;
either the emission region is very optically thin or the resistivity is
dominated by plasma wave scattering, and is far in excess of its independent
particle value.  Either or both of these possibilities is acceptable.

The electric field (Equation 38) may be evaluated, using Equation 34 for
$n_e$ and Equation 35 for $\ell$, and defining the characteristic electric
field $E_c \equiv m_e^2 c^3/(e \hbar) \equiv B_c = 4.4 \times 10^{13}$ cgs:
$$\eqalign{E&\sim \left({p \xi Z \ln\Lambda \over \ell t^{3/2}} {m_e^4 c^5
\over \hbar^3}\right)^{1/2}\cr &\sim {p \alpha^2 \xi^{1/2} \over \tau t^{1/2}}
E_c.\cr} \eqno(44)$$
If the energy release is driven by magnetic reconnection it is proper to use
Equation 43 for $\tau$, yielding
$$E \sim {p \alpha \over b} E_c. \eqno(45)$$
At Galactic halo distances the brightest GRB may have $E$ sufficient to
produce vacuum breakdown into a pair gas (Smith and Epstein 1993), but in
less intense or closer GRB this will not occur, and the resistively heated
ion-electron plasma discussed here may be sufficient.  On a microscopic
level, of course, the power is determined by the magnetic field strength
and configuration, and by the mechanisms of plasma resistivity which drive
reconnection.

The observed non-thermal gamma-ray spectrum of GRB requires the presence of
a nonthermal distribution of ``runaway'' electrons, consistent with the
large $v_{dr}$ and collective processes discussed above.  It may be relevant
that values of $t \sim 1$, consistent with the radiation of the bulk of the
power of GRB at $\sim$ MeV energies, correspond to a maximum in the
conductivity and therefore, under conditions of a fixed potential drop, to a
maximum in the power dissipation.

If $Z \gg 1$, $\sigma_{el}$ increases approximately linearly with $n_+$ in
the range $Z n_i < n_+ < Z^2 n_i$ because of the increasing density of
charge carriers without a corresponding decrease in their scattering length.
This is in contrast to the usual near-independence of density of
$\sigma_{el}$. Pair production thus may provide a natural thermostat at
$t \sim 1$, reducing the power dissipated in regimes at the pair production
threshold by increasing the conductivity under conditions of constant current.
It is evident, of course, that observable cyclotron and annihilation lines
require require much lower values of $t$, and are plausibly produced by
energy and positrons precipitated on the dense surface layers of the neutron
star, cooled by black-body radiation.
\bigskip
\centerline{4. Discussion}
\medskip
The fundamental problem of GRB phenomenology is the apparent inconsistency
between their spatial distribution, which points strongly toward a
cosmological origin, and the spectral features observed in some GRB,
apparently inconsistent with such an origin.  In this paper I have tried to
reconcile these apparently contradictory data by assuming two disjoint
populations of GRB.  Because the argument for a cosmological Population C is
statistical, while the argument for a Galactic Population G is based on the
observation of spectral lines from only a minority of GRB (perhaps from only
a minority of that Population), it is difficult to assign an individual GRB
to either population, unless it shows spectral lines or is identified with
another astronomical object (the SGR of March 5, 1979 is the only good extant
example of such an identification).

Fortunately, the models discussed in this paper predict another potential
distinguishing characteristic.  GRB in Population C produce their radiation
by the synchrotron process of relativistic electrons.  In the magnetic field
is ordered the radiation will be linearly polarized.  GRB in Population G
produce their radiation by the cyclotron process of semi-relativistic
electrons, Coulomb scattered into excited Landau states.  This radiation is
elliptically polarized, with a substantial circular component.  In contrast,
radiation produced by annihilation in a pair gas, such as the initial burst
from a fireball or the trapped pair gas discussed by Carrigan and Katz
(1992), is unpolarized.

The predicted characteristic frequency of gamma-ray emission in fireball
debris impact models (Equation 30) is fairly sensitive to $\gamma_F$.  The
value of $\gamma_F$ depends on physical conditions within the fireball
(Shemi and Piran 1990), and a wide range of $\gamma_F$ would be expected,
with a corresponding range in spectra.  In particular, smaller values of
$\gamma_F$ would lead to X-ray, ultraviolet, or visible bursts, which should
be searched for.  Bursts of lower frequency radiation should have longer
durations and smoother time histories (Equation 28), and perhaps also lower
efficiencies, as accelerated electrons undergo adiabatic expansion before
they radiate their energy.

I thank P. C. Joss and I. A. Smith for discussions.
\vfil
\eject
\def\ref{\medskip \hangindent=20pt \hangafter=1}
\parindent=0pt
\centerline{References}
\ref
Atteia, J.-L., Barat, C., Hurley, K., Niel, M., Vedrenne, G., Evans, W.~D.,
Fenimore, E.~E., Klebesadel, R.~W., Laros, J.~G., Cline. T., Desai, U.,
Teegarden, B., Estulin, I.~V., Zenchenko, V.~M., Kuznetsov, A.~V., and Kurt,
V.~G.~1987, {\it Ap.~J.~(Suppl.)}~{\bf 64}, 305.
\ref
Brainerd, J.~J.~1992, {\it Nature} {\bf 355}, 522.
\ref
Carrigan, B.~J., and Katz, J.~I.~1992, {\it Ap.~J.}~{\bf 399}, 100.
\ref
Cavallo, G., and Rees, M.~J.~1978, {\it Mon.~Not.~Roy.~Astr.~Soc.}~{\bf
183}, 359.
\ref
Cline, T.~L.~1980, {\it Comment.~Ap.}~{\bf 9}, 13.
\ref
Fenimore, E.~E., Epstein, R.~I., Ho, C., Klebesadel, R.~W., and Laros,
J.~1992, {\it Nature} {\bf 357}, 140.
\ref
Fishman, G.~J.~1993, in {\it Proc.~Compton Symp.}, ed.~N.~Gehrels (AIP, New
York) in press.
\ref
Goodman, J.~1986, {\it Ap.~J.~(Lett.)}~{\bf 308}, L47.
\ref
Higdon, J.~C., and Lingenfelter, R.~E. 1990, {\it Ann.~Rev.~Astron.~Ap.}~{\bf
28}, 401.
\ref
Katz, J.~I.~1982, {\it Ap.~J.}~{\bf 260}, 371.
\ref
Katz, J.~I.~1992, {\it Ap.~Sp.~Sci.}~{\bf 197}, 163.
\ref
Katz, J.~I.~1993, in {\it Proc.~Compton Symp.}, ed.~N.~Gehrels (AIP, New York)
in press.
\ref
Landau, L.~D., and Lifshitz, E.~M.~1959, {\it Fluid Mechanics}
(Addison-Wesley, Reading, Mass.).
\ref
Mao, S., and Paczy\'nski, B.~1992, {\it Ap.~J.~(Lett.)}~{\bf 388}, L45.
\ref
Meegan, C.~A., Fishman, G.~J., Wilson, R.~B., Paciesas, W.~S., Pendleton,
G.~N., Horack, J.~M., Brock, M.~N., and Kouvelioutou, C.~1992, {\it Nature},
{\bf 355}, 143.
\ref
M\'esz\'aros, P., and Rees, M.~J.~1993, {\it Ap.~J.}~in press.
\ref
Paczy\'nski, B.~1986, {\it Ap.~J.~(Lett.)}~{\bf 308}, L43.
\ref
Piran, T.~1992, {\it Ap.~J.~(Lett.)}~{\bf 389}, L45.
\ref
Ruderman, M.~A.~1975, {\it Ann.~N.~Y.~Acad.~Sci.}~{\bf 262}, 164.
\ref
Schmidt, W.~K.~H.~1978, {\it Nature} {\bf 271}, 525.
\ref
Shemi, A., and Piran, T.~1990, {\it Ap.~J.~(Lett.)}~{\bf 365}, L55.
\ref
Smith, I.~A., and Epstein, R.~I.~1993 {\it Ap.~J.}~in press.
\ref
Spitzer, L.~1962, {\it Physics of Fully Ionized Gases} (Interscience, New
York).
\ref
Usov, V., and Chibisov, G.~1975, {\it Sov.~Astron.-AJ} {\bf 19}, 115.
\vfil
\eject
\centerline{Figure Captions}
\bigskip
Figure 1: Flow geometry in frame of contact discontinuity CD.  S1 and S2 are
shocks and numbers denote regions of fluid.
\medskip
Figure 2: Emission geometry.
\medskip
Figure 3: Radiation pulse emitted at radius $r$, plotted for $\theta_0^2 =
20c/u$.  Dashed lines denote rise and fall for a cloud sharply circumscribed
between $\theta_1 = (\tau_1 \theta_0^2 - 2c/u)^{1/2}$ and $\theta_2 =
(\tau_2 \theta_0^2 - 2c/u)^{1/2}$.  Note that abrupt rise at $\tau =
2c/(\theta_0^2 u)$ is obtained for a uniformly filled medium, independent of
the angular distribution of radiation, and does not require a cloud bounded in
angle.  Units of the ordinate are arbitrary.
\vfil
\eject
\bye
\end